\documentclass[aps,pre,floatfix,showpacs, superscriptaddress,]{revtex4-2}
\usepackage{amssymb,amsmath,amsfonts}
\usepackage{graphicx,graphics}
\usepackage{epsfig}
\usepackage{epstopdf}
\usepackage{bm,bbm}
\usepackage{mathrsfs}
\usepackage{calc}
\usepackage{float}
\usepackage{color}
\usepackage{ulem}
\usepackage{natbib}

\begin{document}

	\title{Numerical prediction of the steady-state distribution under stochastic resetting from measurements}

\author{Ron Vatash}
\affiliation{Raymond \& Beverly Sackler School of Chemistry, Tel Aviv
	University, Tel Aviv 6997801, Israel}
\author{Amy altshuler}
\affiliation{Raymond \& Beverly Sackler School of Chemistry, Tel Aviv
	University, Tel Aviv 6997801, Israel}
\author{Yael Roichman}
\email{roichman@tauex.tau.ac.il}
\affiliation{Raymond \& Beverly Sackler School of Chemistry, Tel Aviv
	University, Tel Aviv 6997801, Israel}
\affiliation{Raymond \& Beverly Sackler  School of Physics \& Astronomy, Tel Aviv
	University, Tel Aviv 6997801, Israel}
	
	\date{\today}

	\begin{abstract}
	A common and effective method for calculating the steady-state distribution of a process under stochastic resetting is the renewal approach that requires only the knowledge of the reset-free propagator of the underlying process and the resetting time distribution.  The renewal approach is widely used for simple model systems such as a freely diffusing particle with exponentially distributed resetting times. However, in many real-world physical systems, the propagator, the resetting time distribution, or both are not always known beforehand. In this study, we develop a numerical renewal method to determine the steady-state probability distribution of particle positions based on the measured system propagator in the absence of resetting combined with the known or measured resetting time distribution. We apply and validate our method in two distinct systems: one involving interacting particles and the other featuring strong environmental memory. Thus, the renewal approach can be used to predict the steady state under stochastic resetting of any system, provided that the free propagator can be measured and that it undergoes complete resetting.
\end{abstract}

\maketitle
%------------------------------------------------

	\section{Introduction}\label{sec1}

Stochastic resetting (SR) refers to the abrupt random interruption of a dynamical process, followed by its immediate re-initiation \cite{evans11,Evans20}. Repeated resettings of a diffusive process drive it to a nontrivial, out-of-equilibrium, steady state, which is characterized by a stationary density profile \cite{evans11, Evans20,ofir20}. Stochastic resetting has been widely applied to model various physical and natural phenomena \cite{Evans20, gupta2022, Nagar23ManyBodyReset}, including the height distribution of fluctuating interfaces \cite{Gupta2014, Gupta2016} and the thermodynamics governing cell division \cite{Genthon2022}. When resetting occurs at a constant rate with exponentially distributed times, the resulting steady-state density exhibits a Laplacian form determined by the ratio of the resetting rate to the diffusion coefficient \cite{evans11}. The shape of the steady state position distributions depends on the resetting time distribution \cite{Roldan2017}, on the resetting protocol, e.g. partial resetting, where resetting is done only part of the way to the origin \cite{OfirPartialRestart22, MetzlerSoftResettingHarmonicPotential22, Metzler23PartialRestart, Kristian24_PartialResetting}, and on the underlying process, e.g. processes with memory or ballistic motion \cite{BoyerResetwithMemoryPRL14, Majumdar24_PRE_ActiveParticleWithMemory}.
%%%%%%%%%%%%%%%%%%%%%%%%%%%%%%%%%%%%%%%%%%%%%%%%%%%%%%%%%%%%%%%%%%%%%%%%%%%%%%%%%%%%%%%%%%%%%%%%%%%%%%%%%%%%%%%%%%%%%%%%%%%%%%%%%%%%%%%%%%%%%
%% The renewal

One method commonly used to calculate the steady-state position distribution function (PDF) under resetting is solving the renewal equation \cite{evans11}, 
\begin{equation}
	\label{eq:LastRenewalGeneralIntegralForm}
	\rho(x|x_0) = r \int_0^\infty \Psi(t) C(x, t|x_0) dt.
\end{equation}
Here, $x_0$ is the initial position, \(\Psi(t)=  \int_t^\infty p(t)dt\) is the probability of a process to survive without resetting until time $t$, under a resetting time distribution $p(t)$ with a constant mean resetting rate $r$, and \(C(x,t|x_0)\) is the conditional reset-free propagator which satisfies the initial condition \(C(x, t = 0) = \delta(x-x_0) \).
For example, the propagator of a Brownian particle is given by,
\begin{equation}
	\label{eq:DiffEqGaussianSimpleSolutionSingleParticle}
	C(x,t|x_0)=\frac{1}{\sqrt{4\pi D t}}e^{-\frac{(x-x_0)^2}{4Dt}}.  
\end{equation}
where \(D\) is the diffusion coefficient. Under stochastic resetting, with instantaneous returns and exponentially distributed resetting times we obtain for \(t \geq 0\),
\begin{equation}
	\label{eq:SSdistRenewalExponentialSingleParticle}
	\rho(x|x_0) = \sqrt{\frac{r}{4D}} e^{(-\sqrt{\frac{r}{D}} |x - x_0|)} ,
\end{equation}
where \(r\) is the mean resetting rate \cite{evans11}. 

Importantly, the renewal equation indicates that the steady-state distribution of the system can be determined solely from the reset-free propagator and the resetting time distribution without the need for direct measurement. This has significant ramifications for the design of resetting protocols and for the analysis of systems in which the resetting appears naturally. Specifically, in applications such as accelerating state-to-state transitions \cite{Remi24} and search processes \cite{Evans20}, the efficiency depends on the resetting rate \cite{Shlomi16optimalSearch}. Having the ability to evaluate the steady-state distribution at any resetting rate, based only on the knowledge of the free propagator, is highly desirable and is expected to reduce the characterization time significantly. 

However, in realistic and complex systems, the free propagator and the resetting time distributions are not always known in advance. For example, in many-body interacting systems, the free propagator is not readily calculated. The resetting time distribution in natural processes may be governed by the internal properties of the systems and can not be prescribed externally or known a-priory. 

In this paper, we tackle this issue by developing a numerical approach to solve the renewal equation based on measurements. Specifically, we measure the free propagator and, if needed, the resetting time distribution in cases where either the free propagator, the resetting time distribution, or both are unknown. We first validate our method by applying it to a single diffusing colloidal particle for which analytical results are known (Eq.~\ref{eq:DiffEqGaussianSimpleSolutionSingleParticle} and Eq~\ref{eq:SSdistRenewalExponentialSingleParticle}). We then apply our method to two different experimental systems. The first is a many-body system consisting of six colloidal particles undergoing stochastic resetting with holographic optical tweezers (HOT)\cite{SteadyStatePaper_SoonToBePublished}, with exponentially distributed resetting times. We will refer to this system as the \textit{Colloidal system}. The second system is a self-propelled particle moving through an array of mobile obstacles, which it can push aside as it moves. The particle is reset conditionally, either when it reaches the boundaries of the arena or, if it survived, after 20 seconds \cite{Amy24}. We will refer to this system as the \textit{Bug system}. 

\section{Numerical evaluation of the steady-state distribution}
\label{sec:TheMethod}

Experimental measurements are naturally discrete. Therefore, we start our numerical evaluation of the renewal equation by representing the free propagator as a matrix of size \(T\) x \(N\), where \(T\) is the total number of time steps, and \(N\) is the total number of position points at which the propagators are evaluated experimentally, 

\begin{equation}
	\label{eq:MatrixPropagatorForm}
	C(x, t|x_0) \approx \boldsymbol{C_{T \times N}} =  
	\begin{pmatrix}
		C(x_1, t_1) & C(x_2, t_1) &  \cdots & C(x_N, t_1) \\
		C(x_1, t_2) & C(x_2, t_2) & \cdots & C(x_N, t_2) \\
		\vdots & \cdots & \ddots & \cdots \\
		C(x_1, t_T) & C(x_2, t_T) & \cdots & C(x_N, t_T) 
	\end{pmatrix}, 
\end{equation}

\noindent with each element $c_{ij}= C(x = x_j, t = t_i | x_0)$ representing the probability density function value in a given time step \(i\) (\(i \in \{1, T\} \)) and a given position \(x_j\) (\(j \in \{1, N\} \)). 

In analogy to Eq.~\ref{eq:MatrixPropagatorForm}, we can write down the survival function of the resetting times distribution  \(\Psi\) as a row vector of size \(T\) containing the values \(\psi_i= \Psi(t = t_i)\) of every specific time \(t_i\)
\begin{equation}
	\label{eq:PsiRowVec}
	\boldsymbol{\Psi_{1 \times T}} = \left( \Psi(t_1), \Psi(t_2), \cdots, \Psi(t_T) \right).
\end{equation}
This representation allows us to apply the renewal approach numerically by replacing the integral in Eq.~\ref{eq:LastRenewalGeneralIntegralForm} with a right rule Riemann sum over discrete and constant time points \(\Delta t\),
\begin{equation}
	\label{eq:RenewalSumForm}
	\rho(x|x_0) \approx r \sum_{i=1}^{t_{\text{ss}}} \Delta t \Psi( t_i) C(x, t_i| x_0),
\end{equation} 
where \(t_{\text{ss}}\) represents the steady-state time at which the probability density function (PDF) becomes stationary, and \(\Delta t\) is the size of the time-step. 

\begin{figure}[h]
	\centering
	\includegraphics[scale = 0.4]{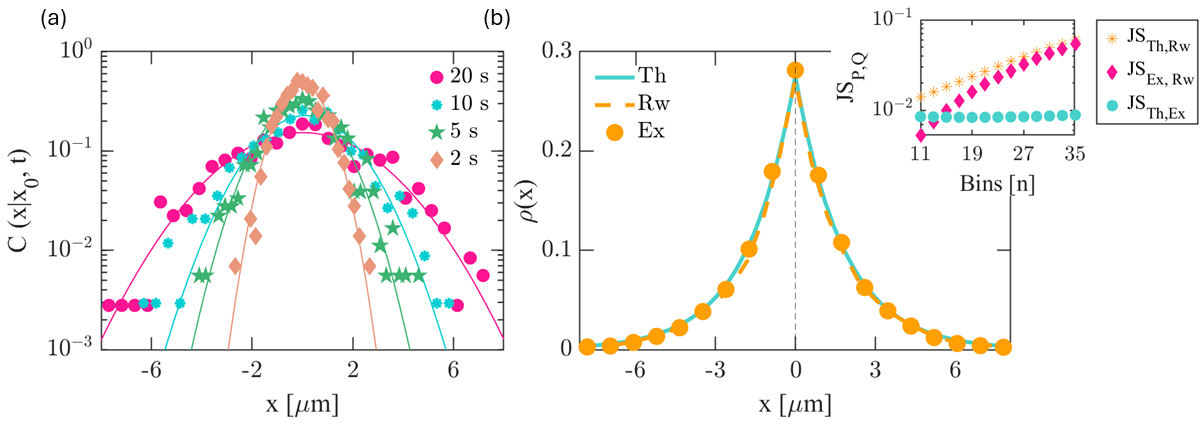}
	\caption{Single particle measurements; (a) Semi-logarithmic curves of the probability density function of a particle's position \(x\) at four different times: 2\thinspace{s} (diamonds), 5\thinspace{s} (stars), 10\thinspace{s} (small circles), and 20\thinspace{s} (large circles). (b) Steady-state distribution comparison of a single particle: analytical solution (solid line, Eq.~\ref{eq:SSdistRenewalExponentialSingleParticle}), numerical renewal approximation (dashed line, Eq.~\ref{eq:RenewlMatForm}), and experimental data (circles). Inset shows the absolute values of Jensen-Shannon divergence \(JS_{P,Q}\) for different pairs of \(P\) and \(Q\): theory vs. renewal (star), experiment vs. renewal (diamond), and theory vs. experiment (circles). The resetting rate was \(r = 0.05 \thinspace{s^{-1}}\).}  
	\label{fig:SingleParticleGraphs}
\end{figure}

Using this approximation, along with the matrix formalism of the reset-free propagator and the survival function of the resetting times, we can now express \(\rho_j\), the probability density function of the process under stochastic resetting evaluated at position \(x_j\) as,
\begin{equation}
	\label{eq:RenewlMatForm}
	\rho_j = \Sigma_i\left( \frac{\Delta t}{\langle T_R \rangle} \right)  \psi_i c_{ij}.
\end{equation}
%The primary source of error in approximating the integral as a sum starting at \(t_1 = 1/\text{fps}\) arises from the first term \(\rho(x_0)\).
%, and reads,  
%\begin{equation}
%   \label{eq:ErrEstimateReiman}
%   \sigma_{\Delta t} \leq \frac{\Delta t}{2} \Bigl[ \Psi(t_0)C(x, t_0|x_0) - \Psi(t_1)C(x, t_1|x_0) \Bigr].
%\end{equation}
%To satisfy normalization and the initial condition $C(x, t = 0|x_0) = \delta(x-x_0)$, we define $C(x, t = 0|x_0) = 1/\Delta x$ in the interval $[-\Delta x/2,\Delta x/2]$ and 0 elsewhere. $\Delta x$ represents the spatial discretization, i.e., bin width.

We apply our method to a single freely diffusing colloidal particle to verify our measurement-based numerical evaluation of the steady-state PDF. The experimental system consists of a dilute suspension of colloidal particles (Silica, 1.5\thinspace{$\mu m$} in diameter) in water, floating slightly above the floor of a glass chamber with $D=0.164 \pm 0.004\thinspace{\mu m^2/s}$. The motion of the particles is captured by a camera (grasshopper 3, point gray) at 30\thinspace{fps}. Conventional particle tracking techniques \cite{CROCKER1996VideoAnalysis} are used to extract the particle's trajectories.

We start by measuring the single particle free propagator and compare it to Eq.~\ref{eq:DiffEqGaussianSimpleSolutionSingleParticle} using the same diffusion coefficient $D$ (Fig.~\ref{fig:SingleParticleGraphs}a). As expected, experiment and theory agree well within measurement error. The experimental error has two main sources, the localization error, which is approximately 30\thinspace{nm}, and the statistical error arising from the finite number of experiments. The latter is most evident at the tails of the distribution, where rare event statistics are required for appropriate sampling (see. Fig.~\ref{fig:SingleParticleGraphs}). Here, a good representation of the Gaussian distribution is obtained for $C(x,t|x_0)>10^{-2}$. 

Next, we implement a resetting protocol with exponentially distributed resetting times, $p(t)=re^{-rt}$ and $r=0.05\thinspace{s^{-1}}$ using home-built automated holographic optical tweezers (HOTs) as described previously \cite{ofir20, SteadyStatePaper_SoonToBePublished}. In short, to reset the particle position, we image the sample, locate the particle, send an optical trap to its position, and drag it back with constant speed to the origin by projecting sequentially a set of partially overlapping optical traps.  To obtain trajectories with instantaneous returns, we cut out the return phase of the trajectories \cite{ofir20,SteadyStatePaper_SoonToBePublished}, leaving only the diffusing phase part. These trajectories are used to obtain the steady-state PDF of the diffusing particle under instantaneous stochastic returns. 

In Fig.~\ref{fig:SingleParticleGraphs}b, we compare the directly measured steady-state PDF (circles) to theory (Eq.~\ref{eq:DiffEqGaussianSimpleSolutionSingleParticle}, solid line) and to its numerical evaluation by our renewal method (Eq.~\ref{eq:RenewalSumForm}, dashed line). This confirms the validity of our numerical renewal scheme.
The error in estimating the real PDF from experiments using the numerical renewal method depends on three main factors: the size of the measured sample, the number of bins used to evaluate the propagator and steady-state PDF (the spatial resolution), and the experimental sampling rate (the temporal resolution). We estimate the sample size error by comparing the steady-state PDF evaluated on 10 different data sets containing half of our sample. We use the Jansen-Shannon divergence $JS$ to quantify the difference between these distributions, where the discrepancy between a reference probability $P$ and a tested one $Q$ is given by, $JS_{P, Q} = 1/2 D_{KL}(P \vert\vert M) + 1/2 D_{KL}(Q \vert\vert M)$, with $D_{KL}(P \vert\vert M) = \sum_{x \in X} P(x)\log(\frac{P(x)}{M(x)})$ the Kullback-Leibler divergence, and $M = 1/2(P + Q)$.
We note that when \(M = 0\), \(P\) is also zero, and the relative entropy vanishes as well, since ($\lim_{x\to 0} x\log(x) = 0 $).

\begin{figure}[t]
	\centering
	\includegraphics[scale = 0.4]{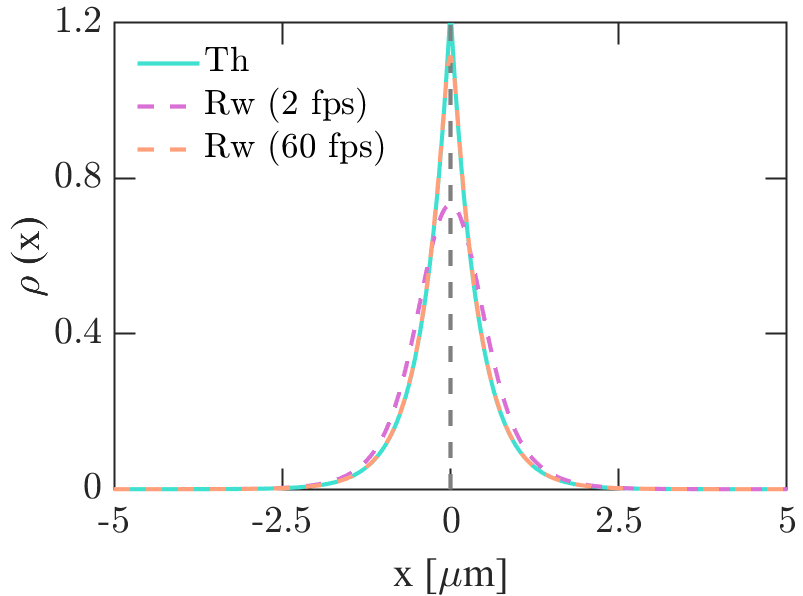}
	\caption{Effect of sampling; A comparison between the integration solution (cyan) to approximation using right Riemann sum in \(2\thinspace{\text{fps}}\) (purple) and \(60\thinspace{\text{fps}}\) (orange) for a single diffusing particle undergoing Poissonian stochastic resetting with resetting rate of $r = 1\thinspace{s^{-1}}$, and diffusion coefficient of $D = 0.164\thinspace{\mu m^2/s}$.}
	\label{fig:effectofsampling}
\end{figure}

The average JS divergence between the different sub-samples (45 pairs) of our numerically evaluated steady-state distributions is $\langle JS\rangle_{sm}= 6\cdot10^{-5}$. We determine the number of bins to maximize precision while retaining a smooth-looking distribution. Finally, we estimate the error stemming from the finite experimental sampling rate by evaluating the calcualetd steady-state PDF under two sampling rates, $2\thinspace{\text{fps}}$ and $60\thinspace{\text{fps}}$.
%$\sigma_{\Delta t}$ from Eq.~\ref{eq:ErrEstimateReiman}, we have $\sigma_{\Delta t} = 2.35$, and $\sigma_{\Delta t} = 0.04$ for $2\thinspace{\text{fps}}$ and $60\thinspace{\text{fps}}$ respectively. Ideally, the sampling rate is fast enough (in comparison to the mean resetting time) that $\sigma_{\Delta t}$ is smaller than the sample error. 
The effect of the sampling error is demonstrated in Fig.~\ref{fig:effectofsampling}, where we compare Eq.~\ref{eq:LastRenewalGeneralIntegralForm} to the right Riemann sum (Eq.~\ref{eq:RenewlMatForm}) for two different values of \(\Delta t\) (\(2\thinspace{\text{fps}}\) and \(60\thinspace{\text{fps}}\)) at resetting rate of $r = 1\thinspace{s^{-1}}$. %and two different resetting protocols. 
%We observe that a low sampling rate results mainly in an oversampling of the distribution at the origin. 
We observe that a low sampling rate results in the absence of a distinct cusp in the distribution. This is due to the resulting significant error in estimating the first term of the evaluated sum. In addition, at low resetting rates, the tails of the distribution are under-sampled, leading to a Gaussian-like distribution.

	Having established an accuracy estimation method and benchmark, we proceed to apply our numerical renewal method to two different experiments. The first consisted of six colloidal particles undergoing stochastic resetting simultaneously, with exponentially distributed resetting times for which the steady-state PDF is a priori unknown. The second consists of a self-propelled bristle bot in an arena of mobile obstacles, for which neither the steady-state PDF nor the resetting time distributions are known in advance. 
	
	%\section{Experimental Model Systems}
	\section{Global stochastic resetting of a colloidal suspension}
	\label{sec:ExpModSysColloids}
	%\textit{Colloidal Suspension under Stochastic Resetting  - Global Resetting with Holographic Optical Tweezers}

	%%%%%%%%%%%%%%%%%%%%%%%%%%%%%%%%% Describe the system %%%%%%%%%%%%%%%%%%%%%%%%%%%%%
	\begin{figure} [h!]
		\centering
		\includegraphics[scale = 0.5]{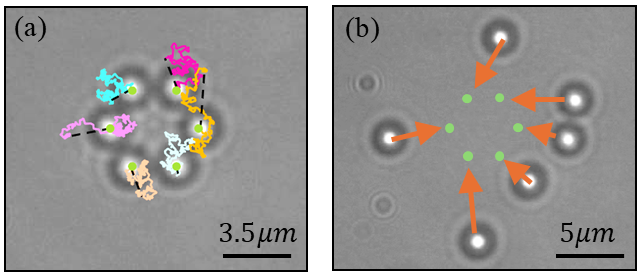}
		\caption{Global resetting of six silica colloidal particles using automated holographic optical tweezers. (a) Particles are initially trapped in their starting arrangement with overlaid typical trajectory segments of free diffusion. Green circles represent the positions of the optical traps, and black dashed lines indicate the return paths. (b) An illustration of the global resetting protocol, where all particles reset simultaneously. Orange arrows depict the assigned trap to which each particle teleports.}
		\label{fig:TrajAndGlobalExplained}
	\end{figure}
	
	To study the stochastic resetting of a many-body system, we perform the following experiments (see also \cite{SteadyStatePaper_SoonToBePublished} for more details). We use our HOTs setup to create six optical traps arranged at the vertices of a hexagon centered around the origin at a distance of \(r_0 = 2.5 \thinspace{\mu m}\). We project these traps on a dilute colloidal suspension in the same configuration described above for the single-particle experiments. 
	
	The experiment starts when six particles are trapped, one at each of the hexagon vertices (Fig~\ref{fig:TrajAndGlobalExplained}a). The traps are then turned off, allowing the particles to diffuse freely for 2 minutes; this constitutes a single-release experiment. Next, the particles are recaptured, and the process starts afresh. Here, we performed 3000 single-release experiments, providing a robust dataset for our analysis. From these experiments, we can generate multiple sequences of stochastic resetting capitulating on the complete memory erasure in a global instantaneous resetting protocol, i.e., a protocol in which the entire system is reset instantaneously (Fig~\ref{fig:TrajAndGlobalExplained}b). Stochastic resetting trajectories are created in the following manner. First, we choose a random sequence of resetting times from an exponential distribution and a random order of our single-release experiments. Each experiment is then cut short according to its matching resetting time. We note that the duration of the single-release experiments sets an upper bound on the resetting time. We then stitch all the appropriately cropped experiments resulting in a continuous, extended resetting experiment.
	
	In these experiments, the two dominant interactions between colloidal particles are hard-core repulsion and hydrodynamic interactions mediated by the presence of the sample walls \cite{colloidaldispersion89, kim06HydroOceanRotne, SokolovRoichmanHydro11, HarelYael14}. An expression for the many-body propagator, i.e., the spreading of the particles in the diffusive phase, is not available and is hard to calculate due to the interactions between the particles. Thus, this experiment serves as a case study for a system where the free propagator is unknown, while the resetting time distribution is known.
	%%%%%%%%%%%%%%%%%%%%%%%%%%%%%%%%%  Measure propagator %%%%%%%%%%%%%%%%%%%%%%%%%%%%
	
	\begin{figure} [t!]
		\centering
		\includegraphics[scale = 0.4]{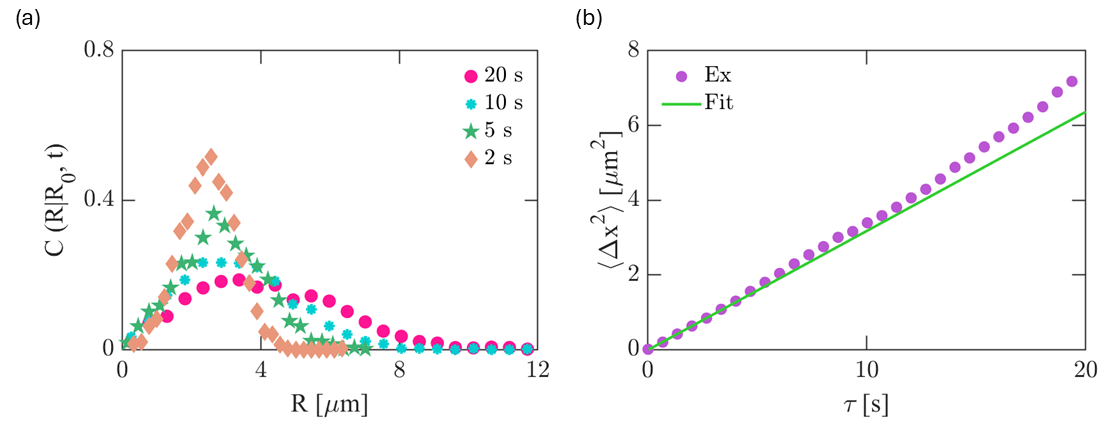}
		\caption{Free propagators of the six colloidal particle system. (a) the probability density function of finding a particle at a radial distance \(R\) at four different times, \(t=2,5,10,20\thinspace{s}\). (b) The ensemble-averaged mean squared displacement (MSD) as a function of lag time (\(\tau\)), circles indicating experimental measurements and the green line representing a linear fit.}
		\label{fig:manybodyFreePropWithMSD}
	\end{figure}
	
	To evaluate the steady-state distribution using the numerical renewal method Eq.\ref{eq:RenewlMatForm}, we first measure the free propagator \(C(x,t)\) by accumulating data from all release experiments. Exploiting the cylindrical symmetry of our experiments, we integrate over the azimuthal direction to obtain the propagator in the radial coordinate $R$. We then discretize time (based on camera capture rate) and radial position to construct the radial discretized probability density matrix ($C$), with elements \(c_{ij} = C(R_i, t_j, | R_0)\). Figure~\ref{fig:manybodyFreePropWithMSD}a displays the free propagators at $t = 2, 5, 10$, and $20\thinspace{}s$ after particle release. 
	
	The ensemble average mean square displacement (MSD) of x-axis projected trajectories is calculated as $\langle \Delta x^2(\tau) \rangle = (1/N_{t}) \sum_{i = 1}^{N_t} \Delta (x_i (\tau))^2$, where $N_t$ represents total trajectories and $\tau$ the lag time. 
	The diffusion coefficient, derived from the MSD slope, shows a gradual increase with time (Fig.~\ref{fig:manybodyFreePropWithMSD}b), which we attribute to changes in effective fluid viscosity. The effective viscosity decreases as particles diffuse apart, consistent with the particle density dependence of fluid viscosity \cite{EinsteinCorrectionVis1911}.
	%%%%%%%%%%%%%%%%%%%%%%%%%%%%%%% Steady-State apply numerical %%%%%%%%%%%%%%%%%%%%%%%%
	\begin{figure} [h!]
		\centering
		\includegraphics[scale = 0.4]{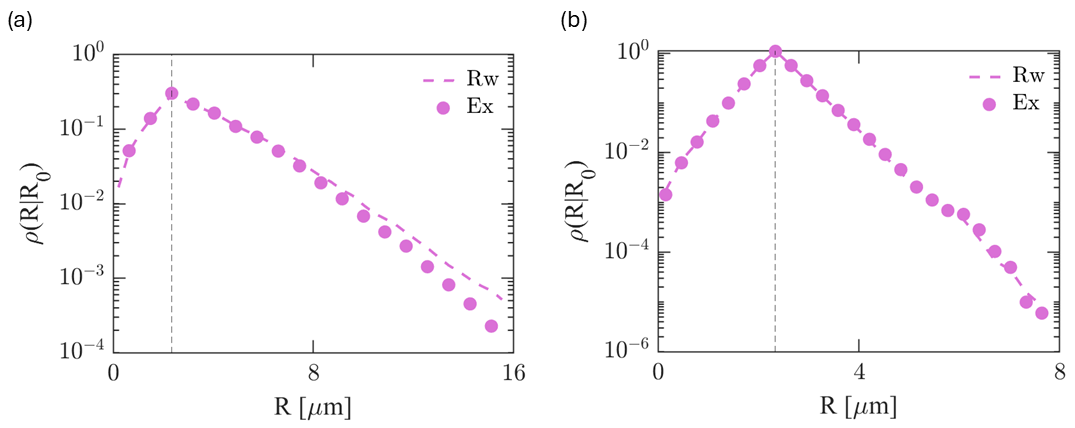}
		\caption{Steady-state distributions of six colloidal particles under global instantaneous stochastic resetting with exponentially distributed resetting times at two different resetting rates on a semi-logarithmic scale: (a) \(r = 0.05\thinspace{s^{-1}}\) and (b) \(r = 1.00\thinspace{s^{-1}}\). Comparison between direct measurements (circles) and predictions from the numerical renewal method Eq.\ref{eq:RenewlMatForm} (dashed line). The initial radial position \(R_0\) is indicated by black dotted lines}
		\label{fig:manybodySteadyStateTheoryVsExp}
	\end{figure}
	
	Following the numerical procedure described in the previous section, we numerically integrate Eq.~\ref{eq:RenewlMatForm} where we have used the measured free propagator $C(R,t| R_0)$ and the Poisson resetting time distribution ($\Psi = e^{-rt})$.
	The resulting numerical prediction for the steady-state PDF of the six-particle system is plotted in Fig.~\ref{fig:manybodySteadyStateTheoryVsExp} (dashed line), for two resetting rates $r=0.05,1.00\thinspace{s^{-1}}$, and is compared to the directly measured one (circles). 
	We find that after \(t_\text{ss} = 60\thinspace{s}\) under stochastic resetting, the colloidal system reaches a steady state in which the accumulated PDF converges. Clearly, the steady-state PDF measured directly and the one predicted by the numerical renewal method agree well within experimental error for all the range at \(r = 1\thinspace{s^{-1}}\) ($JS = 5.0\cdot10^{-5}$), and require more statistics to represent the full range in the challenging conditions of \(r = 0.05\thinspace{s^{-1}}\) ( $JS = 9.4\cdot10^{-4}$).
	%%%%%%%%%%%%%%%%%%%%%%%%%%%%%%%%%%%%%%%%%%%%%%%%%%%%%%%%%%%%%%%%%%%%%%%%%%%%%%%%%%%%%%%%%%%%%%%%%%%%%%%%%%%%%%%%%%%%%%%%%%%
	
	\section{Active Particle with Environmental Memory}
	\label{sec:ExpModSysBug}
	%%%%%%%%%%%%%%%%%%%%%%%%%%%%%%%%%% Describe the system (Bugs) %%%%%%%%%%%%%%%%%%%%%%%%%%%%
	
	\begin{figure} [h]
		\hspace*{-1cm} 
		\centering
		\includegraphics[scale = 0.45]{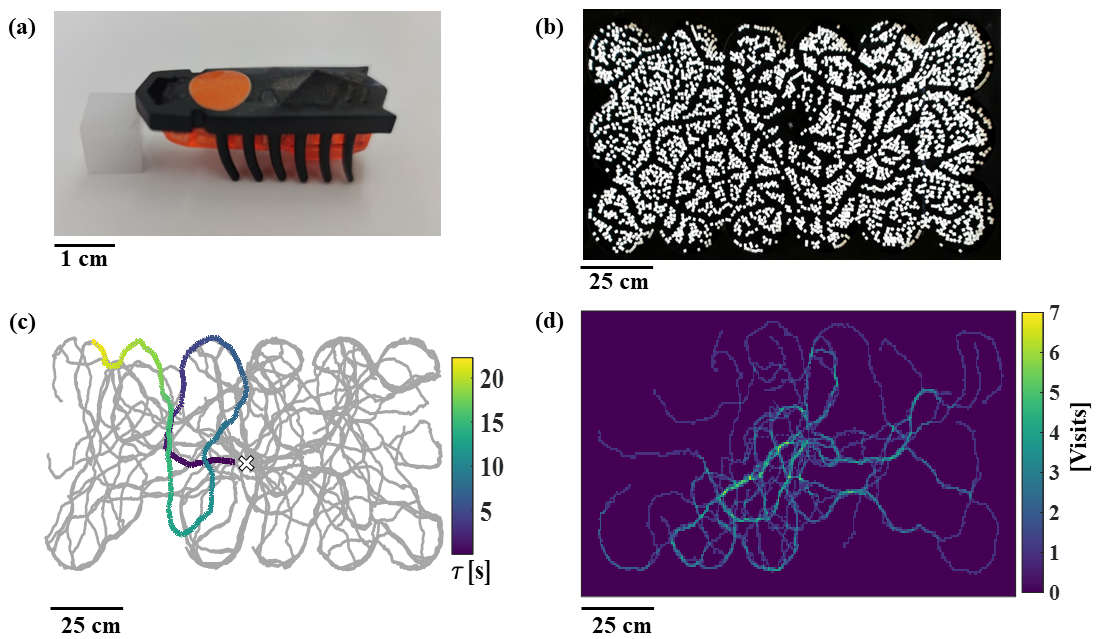}
		\caption{The bug experiment. (a) A single bug and a cubic obstacle, scale bar is $1\thinspace{cm}$. (b) An image of a typical arena structure with environmental memory, $30\thinspace{min}$ into the experiment. (c) Typical trajectories of the bug in a single experiment with 30 resetting events (in gray). The colored trajectory shows the evolution in time of a single trail. A white cross marks the origin. (d) A histogram of the number of visits to different locations in the arena with $0.25\thinspace{cm^2}$ sized bins.}
		\label{fig:Amy_experiment}
	\end{figure}
	
	Building upon the validation of the numerical renewal method in a system with Brownian dynamics and a predetermined resetting time distribution, we apply our method to a second, more challenging experiment.
	This experiment comprises an active bristle robot (Hexbugs, nano, Fig.~\ref{fig:Amy_experiment}a) referred to as "the bug," moving in an arena measuring $90\times150\thinspace{cm}$, filled with cubic obstacles with a side length of $0.95\thinspace{cm}$ (see detailed description in \cite{Amy24}). 
	
	The experiment begins with the bug placed at the arena's center amidst randomly distributed obstacles. As the bug moves, it pushes obstacles aside, creating clear trails that facilitate faster subsequent movement. The system resets every twenty seconds or when the bug reaches the arena's boundaries. During each reset, the bug is manually relocated to the arena's center and rotated 60 degrees counterclockwise from its previous orientation. Importantly, obstacles remain undisturbed during resets, preserving the environmental memory of the bug's past trajectories. The experiment runs for 10 min until reaching a steady state characterized by a dynamic equilibrium of trail formation and destruction (see Fig.~\ref{fig:Amy_experiment}b and \cite{Amy24}).
	
	The bug's motion exhibits characteristics of an active Brownian particle \cite{Dauchot19PRL_DynamicsSelfProp}, with obstacle collisions increasing its rotational diffusion coefficient, as evidenced by the trajectories in Fig.~\ref{fig:Amy_experiment}c. The bug shows a preference for following previously created trails, quantified by the location revisitation probability shown in Fig.~\ref{fig:Amy_experiment}d. This complex behavior, combining active motion with trail-following dynamics, makes it challenging to derive analytical expressions for the free propagator, the steady-state distribution under resetting, and the resetting time distribution.
	\begin{figure} [h] 
		\centering
		\includegraphics[scale = 0.4]{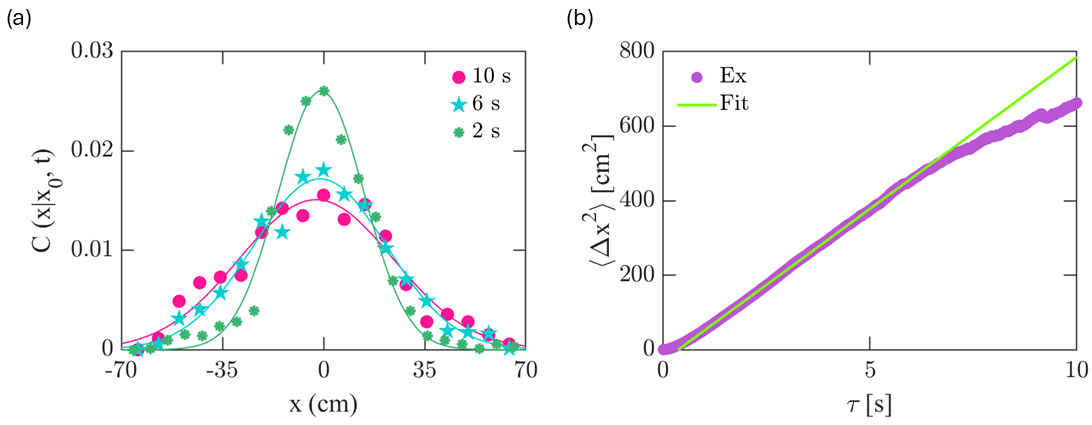}
		\caption{Free propagators of the bug experiment. (a) the probability density function of finding the bug at position $x$ at three different times, \(t=2, 6, 10\thinspace{s}\). (b) the MSD as a function of the lag time \(\tau\). Inset is shown diffusion coefficient as a function of lag-time ($\tau$).}
		\label{fig:BugFreePropMSD}
	\end{figure}
	
	To determine the steady-state distribution using the numerical renewal method, we begin by calculating the free propagator matrix \(C(x,t| x_0)\), focusing on the arena's long dimension where boundary effects are minimized. Figure~\ref{fig:BugFreePropMSD}a shows the experimentally measured free propagators at $t = 2$, $6$ and $10\thinspace{s}$.   While the propagator exhibits a Gaussian-like shape, it develops distinctive features over time, as evident when comparing the distributions at $t=2\thinspace{s}$ to that of $t=6\thinspace{s}$, and $10\thinspace{s}$.
	The origin of these features - whether they reflect statistical fluctuations or environmental memory effects - remains uncertain. The propagator's asymmetry, for instance, might result from the bug's chiral motion. Given this uncertainty, we pursue two approaches for the renewal calculation: first, using a Gaussian fit to the propagators as input for Eq.~\ref{eq:RenewlMatForm}, and second, employing the raw measured propagator data.

	Effects of our reset protocol, which activates when the bug reaches arena boundaries or when $20\thinspace{s}$ elapsed, appear in the MSD curve around $\tau=20\thinspace{s}$. Thus, we truncate the curve presented in Fig.~\ref{fig:BugFreePropMSD}b accordingly.  The nonlinearity observed in the MSD curve at short timescales is likely due to limited positioning precision and the persistent motion of the bug. At intermediate times, the MSD shows linear behavior, while at longer times, a more complex time dependence emerges. This latter behavior reveals the system's underlying memory effects and strong correlations between the bug's successive movements, distinguishing it from simple diffusive dynamics.
	
	%The inset in Figure.~\ref{fig:BugFreePropMSD}b displays the diffusion coefficient of the bug's system as a function of lag-time ($D(\tau)$), calculated using the local numerical derivative ($D_i = (\langle\Delta x^2_{i+1}\rangle - \langle\Delta x_i^2\rangle)/2(\tau_{i+1}-\tau_i$)) and a one-second running average. 

\begin{figure} [t] 
		\centering
		\includegraphics[scale = 0.4]{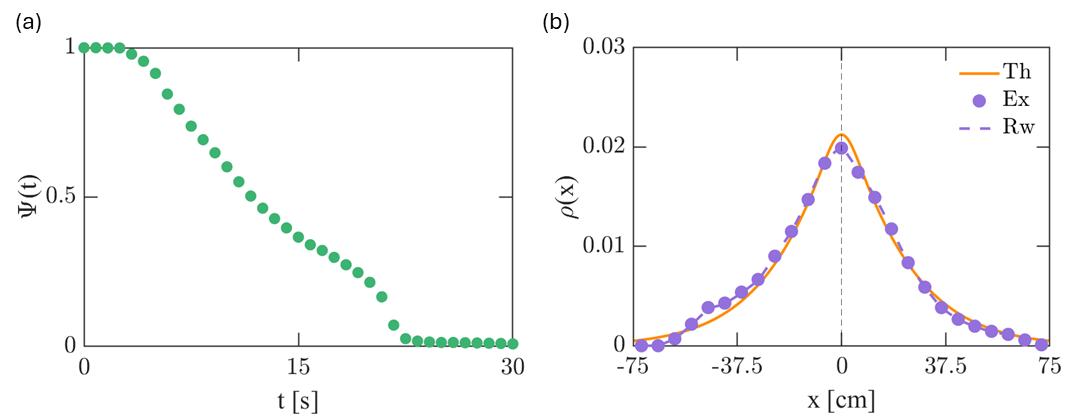}
		\caption{Steady state distribution under resetting of the bug's position. (a) The measured survival probability of a bug integrated numerically from the measured resetting time distribution. (b) Comparison between the measured steady-state distribution (circles) to the numerical renewal calculation with a Gaussian fit to the propagator (solid line) and the directly measured propagator (dashed line)}
		\label{fig:SSdistAndSurvivalBug}
	\end{figure}

	We extract the resetting time probability from the bug's trajectories and numerically integrate it over time to determine the survival probability, $\Psi(t)$ (Fig~\ref{fig:SSdistAndSurvivalBug}a). Substituting the measured $C(x,t|x_o)$ and  $\Psi(t)$ in Eq.~\ref{eq:RenewalSumForm}, we obtain the renewal-based evaluation for the steady-state distribution under resetting. Figure~\ref{fig:SSdistAndSurvivalBug}b compares three versions of the steady-state distribution: the experimental measurements (circles) and two numerical renewal method predictions - one using a Gaussian fit to the measured propagator with an effective diffusion coefficient $D = 42.7\thinspace{cm^2/s}$ (See~\cite{Amy24}), and another using the raw propagator data. The two numerical predictions yield distinct steady-state distributions. The Gaussian-based calculation, which smooths over features in the measured propagator, fails to capture fine details present in the experimental steady-state distribution. While this smoothing might beneficially eliminate statistical noise, it could also suppress genuine system complexities. However, the calculation using the raw propagator data better reproduces the experimentally observed steady-state distribution with $JS = 3\cdot10^{-5}$. From a practical perspective, this improved agreement is valuable as it better predicts the system's actual behavior, regardless of whether the captured features arise from fundamental dynamics or statistical fluctuations.
	
	%%%%%%%%%%%%%%%%%%%%%%%%%%%%%%%%%%%%% talk about difference in resulting steady-state %%%%%%%%%%%%%%

	%\textcolor{red}{Figure 6a} shows the distribution of resetting times obtained from the lab experiments, while \textcolor{red}{Figure 6b} illustrates the survival probability calculated from this distribution. Finally, \textcolor{red}{Figure 6c} compares the steady state distribution measured in the lab (dots) with two steady-state distributions computed using the numerical renewal method: one where \(\psi\) is based in the exact expression for sharp resetting (Eq.~\ref{eq:PsiSharp}), with \(\langle T_r \rangle = t_\text{sharp} = 20\thinspace{s}\) and the other where \(\psi\) is derived from the lab-measured resetting distribution shown in \textcolor{red}{Figure 6a,b}. For lab measured \(psi(t)\) (\textcolor{red}{Fig}) we obtained a mean resetting time of \(\langle T_r \rangle = 13.04\pm 0.02\thinspace{s} \).

	%we first need to measure the free propagators at every time step, from the first frame to the last, for each of our free diffusion videos (without resetting). Given that our system exhibits cylindrical symmetry with no azimuthal dependence, it is reasonable to conduct the analysis in the radial (\(R\)) coordinate. Additionally, to apply these propagators in equation 2, they must all be evaluated at the same r values. To achieve this, we calculated the propagators within predefined radial segments (r1, r2, etc.). As a result, we now have the probability density function (PDF) for each distance r1 at every time t1.

	\section{Conclusions}
	\label{sec:Conclusions}
	
	In this paper, we presented a numerical approach for predicting the steady-state distribution of systems with unknown reset-free distributions by applying a discrete form of the renewal approach combined with lab-measured free propagators.
	We provided a practical method for predicting the steady-state distribution under stochastic resetting across various resetting rates, eliminating the need for experimental trials. This approach can be especially useful in selecting an optimal resetting rate before conducting experiments. 
	
	This work demonstrates the equivalence and differences between the integral form and matrix form of the renewal approach when the sampling rate is sufficiently high. We discussed the impact of the sampling rate on the accuracy of the predictions.  Moreover, we have shown that steady-state distributions can be accurately obtained, even in the absence of analytical expressions for the underlying process (without resetting), by combining lab measurements and a high sampling rate with the matrix form of the renewal approach.
	
	Our method broadly applies to systems that exhibit interactions, memory effects, or any other complexities as long as the reset-free propagators and resetting time distribution are measurable and the renewal condition applies—i.e., the system fully resets at each resetting event (global resetting). We note that this condition can be relaxed, as shown in the bug system experiments, where the renewal condition applies only to the ensemble average level. Nevertheless, our numerical prediction of the steady-state distribution perfectly matches the experimental measurements, demonstrating the robustness and utility of our method for a wide range of real-world physical systems. 
	
	Finally, we stress that the numerical method more accurately captured distinct features of the system, which were smoothed out and overlooked in the analytical Gaussian-based approximation. This advantage makes our numerical approach a more accurate experimental tool for approximating the steady-state distribution, as it consistently captures the system's physical and statistical characteristics in full detail.

%------------------------------------------------
\begin{acknowledgments}
	This research has been supported by the Israel Science
	Foundation (Grants No.\ 385/21).
\end{acknowledgments}

%------------------------------------------------
% References
%------------------------------------------------

%\bibliography{sn-bibliography}

\providecommand{\noopsort}[1]{}\providecommand{\singleletter}[1]{#1}%

\end{document}